\documentstyle[epsfig]{aipproc}

\begin{document}
\title{RXTE Observations of GX 3+1}

\author{Tod E. Strohmayer}
\address{Laboratory for High Energy Astrophysics, NASA/GSFC, Greenbelt, MD
20771\thanks{USRA research scientist}}
\maketitle
\begin{abstract}

We report on new RXTE observations to search for millisecond X-ray variability
from the bright Galactic bulge X-ray source GX 3+1. Although kilohertz
quasiperiodic oscillations (QPO) have been detected now in 14 low mass X-ray
binary (LMXB) systems they have, somewhat mysteriously, not yet been seen in the
bright, persistent, Galactic bulge sources; GX 3+1, GX 9+1, GX 9+9 and GX 13+1.
These systems have been typically classified as atoll sources and when bright
are seen on the banana or  upper banana branch in the X-ray color - color
diagram. During our observations the source was also in a banana state and we
did not detect kHz QPO. We place an upper limit on the amplitude of kHz QPO of
about 1\% rms. 

\end{abstract}

\section*{Introduction}

GX 3+1 is one of the brightest, persistent Galactic bulge X-ray sources. The
source has occasionally produced thermonuclear X-ray bursts when its X-ray flux
was low (see \cite{Makish,Sun}). Based on X-ray spectral and temporal
variability studies with EXOSAT it has been classified as an atoll
source. From this and the observed type I X-ray bursts the source is almost
certainly a low mass binary containing a neutron star \cite{HVK}.
\cite{Lew87} reported the detection of low frequency noise (LFN) and QPO with
frequencies around 8 Hz in some data intervals using EXOSAT observations.
Subsequent analysis reinterpreted the power spectra of GX 3+1 in terms of so
called very low frequency noise (VLFN) and a ``peaked'' high frequency noise
(HFN) component, which then led to the atoll classification \cite{HVK}.

To date kilohertz quasiperiodic oscillations (kHz QPO) have been detected wth
RXTE in 13 LMXB (see \cite{vdk97} for a recent review).
So far there is no strong consensus on the mechanism and/or mechanisms which
produce the kHz QPO in the persistent X-ray flux from LMXB, but considerable
interest has been focused on various beat-frequency interpretations of the kind
first proposed to explain the twin QPO peaks observed in 4U 1728-34
\cite{Stroh96}. The millisecond timescale of the variability indicates that
whatever process is at work is almost assuredly ocurring within a few 10s of
km of the neutron star surface. Here we present results from an analysis of a
short RXTE observation of GX 3+1 to search for high frequency variability.

\begin{figure}[b!] 
\centerline{\epsfig{file=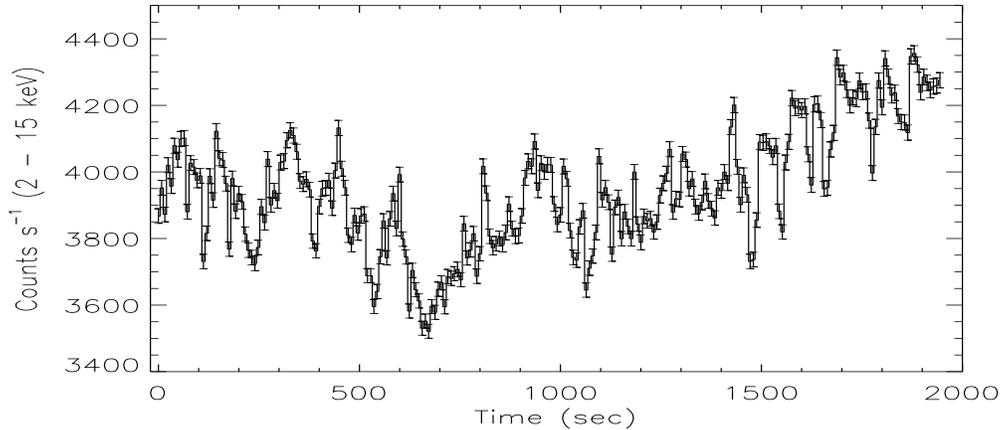,height=2.5in,width=5.5in}}
\vspace{10pt}
\caption{Lightcurve from GX 3+1 measured with the PCA in the 2 - 15 keV range.
The data show the countrates sampled in 8 s time bins. Notice the presence of 
significant non-Poissonian variations on timescales $> 10$ s.}
\label{fig1}
\end{figure}

\section*{Data Summary}

The observations were conducted with RXTE on October 9, 1996 UT.
We obtained 2 ksec of data in 4, 125 $\mu$s (1/8192 s) single bit data modes.
These proportional counter array (PCA) data modes covered photon energies from 
2 - 5, 5 - 9, 9 - 14, and 14 - 90 keV. We also obtained so called Standard modes
which provide 1/8 s lightcurves across the full PCA bandpass as well as 16 s
spectral accumulations. The PCA lightcurve from our observations in the 2 - 15
keV range is shown in figure 1. The data shows the countrates
measured in 8 s intervals. The presence of strong low-frequency variability on
timescales longer than about 10 s is clearly evident in this lightcurve. This
type of strong LFN from GX 3+1 was also reported by \cite{Lew87}.

\begin{figure}[b!] 
\centerline{\epsfig{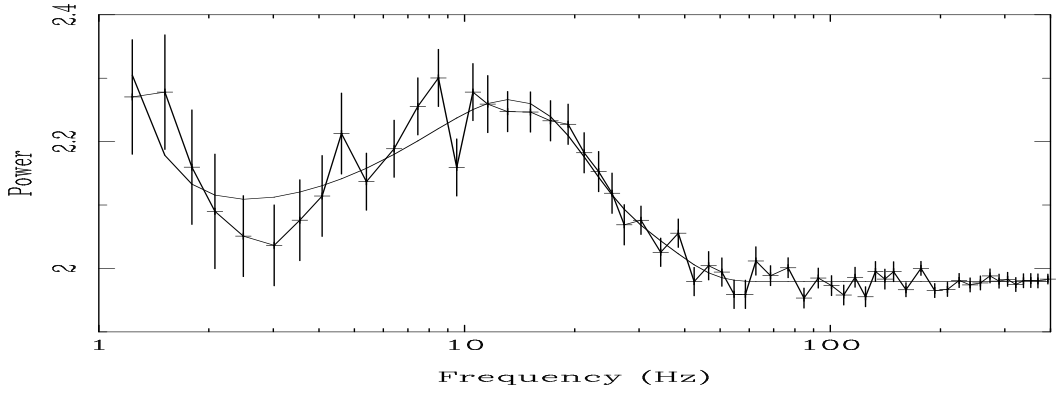}}
\vspace{10pt}
\caption{Power spectrum of GX 3+1 in the 2 - 15 keV range measured with the PCA.
The broad feature between 3 and 30 Hz is the peaked HFN component. The solid
line is the best fitting model described in the text.}
\label{fig2}
\end{figure}

\begin{figure}[b!] 
\centerline{\epsfig{file=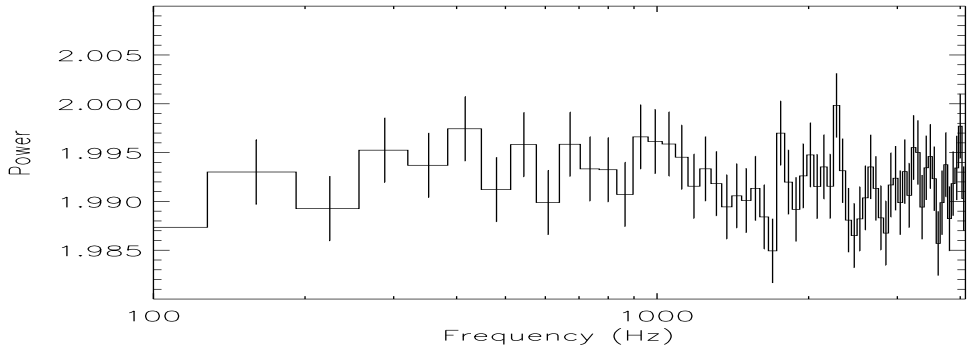,height=2.0in,width=5.5in}}
\vspace{10pt}
\caption{High frequency power spectrum from GX 3+1. There is no significant 
detection of kHz QPO during this observation. The upper limit on the amplitude
of Khz QPO in the 900 Hz region is about 1\% (rms).}
\end{figure}

\section*{Power Spectral Analysis}

We used the single bit data in the 2 - 15 keV range to calculate FFT power
spectra. We calculated power spectra for individual, 128 s intervals and then
averaged the individual spectra. To further improve the power spectral 
statistics we also averaged again in frequency space. Figure 2 shows the power
spectrum from 0.5 to 400 Hz calculated in this way. The broad feature
centered around 10 keV has been referred to as "peaked" high frequency
noise (HFN \cite{HVK}) and also QPO \cite{Lew87}. To model this noise 
component and estimate its amplitude we fit the power spectrum to a model
consisting of a power law and a gaussian. This model provides a marginally
acceptable fit to the data, although it does not fit the shape of the
feature very well at the lower frequencies. From this fit we derive an rms
amplitude for the peaked HFN of about 4 \%, which is similar to values measured
previously with EXOSAT \cite{Lew87}. We also fit a power law model to the 
power spectrum for frequencies below 2 Hz and find a good fit with
$P \propto \nu^{-1.23}$, where $P$ and $\nu$ are the power spectral amplitude
and the frequency, respectively. The amplitude (rms) in the 2 - 15 keV band of
this LFN  is 4.2 \% (integrated from 0.001 to 4 Hz). 
We also investigated the dependence of rms amplitude on photon energy and found
that the amplitude of the LFN increases with photon energy in a way similar
to that reported by \cite{Lew87}.

The power spectrum in the 2 - 15 keV range, emphasizing the high frequency
portion, is shown in figure 3. We did not find any significant QPO in the 
kHz range. There is a modest enhancement in the vicinity of 900 Hz, but it is
not significant enough to claim a detection. The inferred rms
amplitude assuming that all power in that frequency bin were signal power is
about 1 \%. 

\section*{Discussion}

Based on the presence of strong LFN and the peaked HFN the source was almost
certainly in the so called ``banana'' state during these observations.
Whether or not the source is in an upper or lower banana state is difficult to
discern from these short observations.  The lack of kHz QPO is interesting,
since other atoll sources with strong kHz QPO in the island states, such as
4U 1728-34, do not show kHz QPO (or it is very much weaker) when they are in 
the banana state. Indeed, recent observations of 4U 1728-34 in the upper banana
state did not show kHz QPO down to rms amplitudes $< 1\%$ \cite{Stroh97}.
Similar behavior has been reported for 4U 1820-30, which only shows kHz QPO 
in a rather narrow range of source intensities \cite{Smale}. These 
nondetections at high inferred mass accretion rates are an important clue as 
to the mechanism which produces the kHz QPO. In some models the optical 
depth to X-ray photons in the vicinity of the inner edge of the accretion disk
is an important parameter in determining the frequency and strength of kHz
variations \cite{MLS}. For example, it is likely that the optical depth in the
inner accretion disk region increases with mass accretion rate as sources
move from the banana to upper banana branches in the color - color diagram. 
One possibility is that any kHz variability is scattered to much lower 
amplitudes by the increasing optical depth.

\end{document}